# Refining hydrogen positions in α-FeOOH through combined neutron diffraction and computational techniques

Yusuke Nambu, *[a,b] Akihide Kuwabara, [c] Masahiro Kawamata, [d] Seira Mori,[e] Megumi Okazaki [e] and Kazuhiko Maeda [e,f]

The hydrogen positions and magnetic structure of goethite α-FeOOH, a key component of iron rust, were examined through neutron diffraction. All symmetry-allowed magnetic structures under the space group $Pnma$ with the magnetic wavevector $q_m$ = (0, 0, 0) r.l.u. were analysed using irreducible representation and magnetic space group approaches. The magnetic moments aligned along the $b$-axis form antiferromagnetic spin arrangements, as reproduced by first-principles calculations. Accurately determining the hydrogen positions is crucial for understanding the mechanism of catalytic reduction of $CO_2$ in α-FeOOH. These positions were precisely identified through diffraction and calculations, highlighting the effectiveness of using both methods for undeuterated compounds.

## Introduction

Hydrogen is an element of immense significance in compounds, adopting various bonding configurations and structural roles.[1] Its small atomic radius allows it to incorporate into the crystal lattice of many inorganic materials, significantly altering their physical and chemical properties. Understanding hydrogen's incorporation is thus crucial for developing new functionalities and enhancing existing properties, leading to diverse applications in materials science.

One of the most intriguing applications of hydrogen is in the field of high-pressure superconductivity,[2] where compounds like hydrogen sulphide exhibit superconducting properties at remarkably high temperatures.[3] These discoveries enable the development of new superconducting materials with significant technological implications.

Moreover, hydrogen's role in energy production, such as in fuel cells and hydrogen storage systems, highlights its potential for addressing global energy challenges.[4–6] Its participation in catalytic reactions makes it indispensable for numerous industrial processes.[7–9] Hydrogen's ubiquitous nature and unique chemical properties make it a cornerstone of both basic and applied research. It also plays a crucial role in photocatalytic processes, vital for sustainable energy conversion and environmental remediation.[10] These applications underscore the diverse and transformative potential of hydrogen in advancing technology and sustainability.

Goethite (α-FeOOH) is one of the most widely occurring iron oxyhydroxide minerals and is also a main constituent of iron rust,[11,12] and its crystal structure,[13] magnetic properties[14] and surface chemistry have been extensively investigated in mineralogy, solid-state chemistry, and environmental science. Recent studies have shown that α-FeOOH acts as a catalyst for the photochemical reduction of $CO_2$ to HCOOH, in the presence of $[Ru(bpy)_3]^{2+}$ (bpy = 2,2′-bipyridine) as a redox photosensitiser.[15] The rate of HCOOH generation and its selectivity with α-FeOOH are significantly higher than those observed with α-$Fe_2O_3$, indicating an important role for the hydroxyl ($OH^-$) groups in α-FeOOH.[16] Additionally, α-FeOOH exhibits superior $CO_2$ adsorption capabilities compared to α-$Fe_2O_3$.[15,17] Both experimental and theoretical studies over decades have investigated $CO_2$ adsorption on α-FeOOH,[17–19] highlighting the pivotal role of surface OH groups in this process. It is further suggested that these OH groups are involved in proton-coupled electron transfer reactions, which are essential for the reduction of $CO_2$.[20] More concretely, the reduction of $CO_2$ progresses as electrons and protons react with the $CO_2$ adsorbed on the catalyst surface. Therefore, OH groups with proton-dissociation capability on the catalyst surface may play a significant role. In this context, accurate bulk OH geometry and hydrogen-bond topology constrain the local proton environment and provide consistent boundary conditions for modelling surface terminations and proton mobility/transfer steps that may govern catalytic selectivity and rates. Therefore, improving confidence in the bulk H positions and hydrogen-bond geometry is valuable even when the catalytic reaction occurs at surfaces.

[a]*Institute for Integrated Radiation and Nuclear Science, Kyoto University, Kumatori, Osaka 590-0494, Japan. E-mail: nambu.yusuke.7s@kyoto-u.ac.jp*
[b]*FOREST, Japan Science and Technology Agency, Kawaguchi, Saitama 332-0012, Japan*
[c]*Nanostructures Research Laboratory, Japan Fine Ceramics Center, Nagoya, Aichi 456-8587, Japan*
[d]*Department of Physics, Tokyo Metropolitan University, Hachioji, Tokyo 156-0057, Japan*
[e]*Department of Chemistry, School of Science, Institute of Science Tokyo, Tokyo 152-8550, Japan*
[f]*Research Center for Autonomous Systems Materialogy (ASMat), Institute of Science Tokyo, Kanagawa 226-8501, Japan*

Precise determination of hydrogen atom positions within a compound is crucial for materials characterisation. Despite its widespread use in structural analysis, X-ray diffraction struggles to locate hydrogen atoms due to their low electron density, which renders them nearly invisible to X-ray scattering, resulting in incomplete structural information.

Neutron diffraction, on the other hand, is a powerful tool for determining hydrogen positions, overcoming the limitations of X-ray diffraction. Unlike X-rays, neutrons interact with atomic nuclei, making them highly sensitive to hydrogen. This sensitivity allows for precise mapping of hydrogen locations. Recent studies, including single-crystal work, show that accurate hydrogen positioning can be achieved without using polarised neutrons and/or deuteration. Techniques using unpolarised neutrons and light hydrogen provide reliable results, simplifying experimental procedures and broadening the applicability of neutron diffraction in various research fields.

Nuclear magnetic resonance is another tool for locating hydrogen,[21] but it may be less suitable for precise structural refinements of inorganic compounds. Inorganic compound analysis can be challenging due to the influence of heavier atoms and their distribution in crystals. Despite advancements in experimental techniques, first-principles calculations remain essential for a comprehensive understanding of hydrogen in materials.[22] These calculations offer detailed insights into electronic structures, bonding interactions, and dynamical properties that are not easily accessible through experiments alone.

In this study, we examine the hydrogen positions and magnetic structure of goethite α-FeOOH [structures schematically depicted in Fig. 1]. The previous study[13] on α-FeOOH refined hydrogen positions using standard Rietveld analysis. However, ambiguity in the hydrogen locations persists when solely determined by neutron diffraction measurements. Here, neutron diffraction techniques were reapplied to a powder sample, and the magnetic structure was successfully refined using group theoretical analyses. We show that combining complementary first-principles calculations provides an excellent example of locating hydrogen positions in α-FeOOH.

## Methodology

Commercial α-FeOOH powder (Kojundo Chemical Lab. Co., Ltd, Japan, 99%+) was used for neutron diffraction measurements. Neutron powder diffraction data were collected using the high-resolution HERMES diffractometer[23] at the Japan Research Reactor, with a wavelength of $\lambda$ = 1.34239(7) Å. Diffraction patterns were obtained within a temperature range of $T$ = 60 to 485 K using a cryofurnace. Reported temperature uncertainties correspond to $1\sigma$ values obtained from the time-series standard deviation of the calibrated thermometer during the acquisition window. Rietveld refinements were performed using the FullProf Suite,[24] employing a pseudo-Voigt peak-profile function and a Chebyshev polynomial background.

First-principles calculations were performed using the projector augmented-wave method,[25,26] as implemented in the VASP code.[27–29] The exchange–correlation interactions of the electrons were treated within the framework of the PBEsol-type potential.[30] The cutoff energy for the plane wave basis sets was set to 550 eV. The valence-electron configurations of the potentials were $3p^63d^64s^2$ for iron, $2s^22p^4$ for oxygen, and $1s^1$ for hydrogen. All calculations were performed in a spin-polarised state with both ferromagnetic and antiferromagnetic configurations. The antiferromagnetic configurations were investigated in conventional unit cells and double-volume supercells of α-FeOOH. Symmetrically independent antiferromagnetic configurations were identified using the CLUPAN code.[31] Three configurations were calculated in the conventional unit cell and 55 in the doubled-volume supercells.

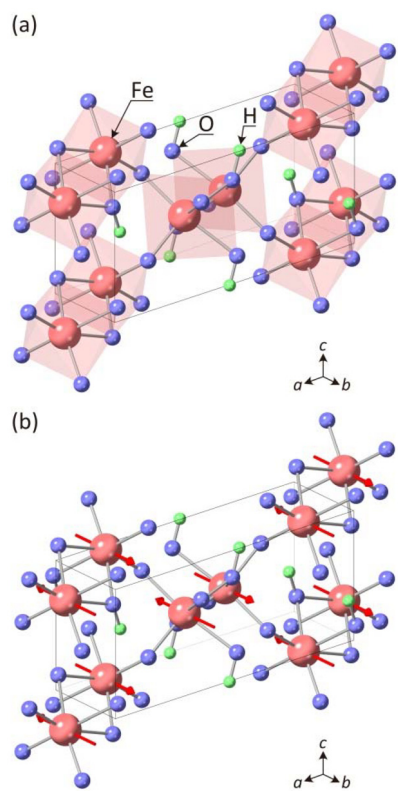

**Fig. 1** (a) Crystallographic unit cell of α-FeOOH with arrows indicating each element. (b) The refined magnetic structure of α-FeOOH, where the magnetic unit cell matches the crystallographic unit cell, reflecting the magnetic wavevector, $\boldsymbol{q}_m$ = (0, 0, 0) r.l.u.

The DFT+U approach[32] was employed to take into account the strong correlation effects of the Fe 3d orbitals, with U set at 5.3 eV.[33] The calculation cells of α-FeOOH were optimised using Γ-centered $k$-point sampling meshes. The $k$-point spacing was set to 0.4 Å$^{-1}$, ensuring total energy convergence to 1 meV per atom against the $k$-points density in the Brillouin zone of the calculated cells. The crystal structures were fully relaxed until all residual forces on the atoms were smaller than 0.02 eV Å$^{-1}$.

# Results and discussion

First, to determine the magnetic transition temperature, we performed powder neutron diffraction as a function of temperature. Fig. 2(a) depicts the raw data collected at $T$ = 64.8(1) and 430.1(9) K, where significant background reflecting the large cross-section of incoherent-scattering of hydrogen (80.26 barn (ref. 34)) is visible. Consistent with the previous studies,[22,35] the crystal structure [Fig. 1(a)] is accounted for by the orthorhombic space group $Pnma$. The structure consists of edge-shared FeO$_6$ octahedra with two sorts of anions, O$^{2-}$ and OH$^-$, and its magnetism is attributed to the high-spin state of Fe$^{3+}$ (3d$^5$) spins, $S$ = 5/2. In addition, an enhancement in neutron-diffraction intensity for several peaks is observed at

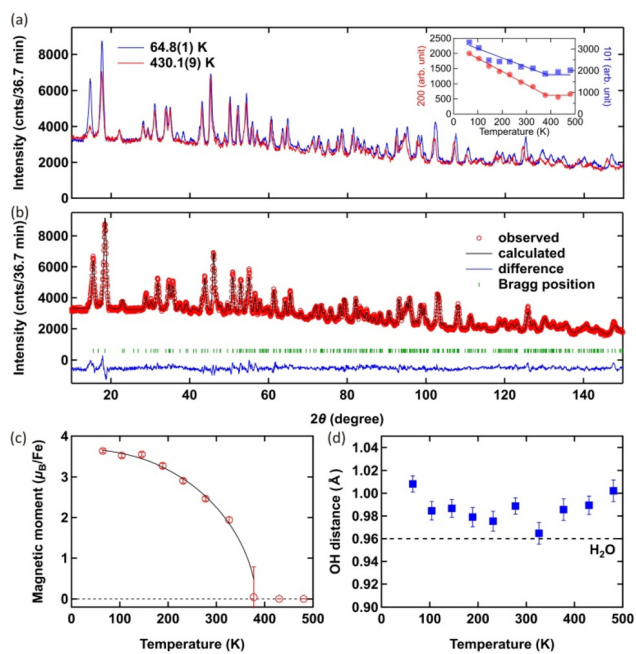

**Fig. 2** (a) Raw neutron powder diffraction data for α-FeOOH collected at 64.8(1) and 430.1(9) K. The inset shows the temperature dependence of the integrated intensity of the 200 and 101 peaks. (b) Neutron powder diffraction data for 64.8(1) K with Rietveld refinement (solid lines). The calculated positions of nuclear and magnetic reflections are indicated by green ticks. The bottom lines show the difference between observed and calculated intensities. Temperature dependence of (c) the evaluated magnetic moment size and (d) the OH distance, with the horizontal dashed line indicating 0.96 Å in H$_2$O.

the lower temperature. The inset to Fig. 2(a) summarises the temperature dependence of the integrated intensity of the 200 and 101 peaks, estimating the Néel temperature, $T_N$, to be 382 (2) K from fits. This value is close to previously reported Néel temperatures for goethite.[13,14,35]

We then applied representation analysis to identify the magnetic structure at the lowest measured temperature. Given that all the magnetic reflections overlap with nuclear reflections, the magnetic wavevector is $\bm{q}_m$ = (0, 0, 0) r.l.u. Basis vectors (BVs) of the irreducible representations (irreps) for the wavevector with the Kovalev notation[36,37] are summarised in Table 1 alongside schematic drawings of magnetic structures corresponding to each BV. There are 12 BVs in total, each belonging to one of 8 distinct one-dimensional irreps. Each BV describes the relation of the moment direction of 4 atoms, either parallel or antiparallel, along one crystallographic axis. The $\psi_4$ in $\Gamma_3$, $\psi_7$ in $\Gamma_5$, and $\psi_{11}$ in $\Gamma_7$ represent ferromagnetic spin arrangements along the $a$-, $b$-, and $c$-axis, respectively, whereas the others describe antiferromagnetic spin arrangements.

We also summarise all possible magnetic space groups (MSGs). Allowed maximal MSGs for the space group $Pnma$ with the wavevector (0, 0, 0) r.l.u. are 8 in total. These MSGs, along with the unified (UNI) symbols[38] and corresponding irreps, are presented in Table 2. The 8 MSGs apparently have a one-to-one correspondence with the irreps.

We evaluated all 8 possible MSGs by comparing the reliable factor $R_{mag}$ from data collected at 64.8(1) K. The best fit was found to be $R_{mag}$ = 8.53% with $Pnma'$ (#62.445), corresponding to $\psi_{12}$ in $\Gamma_8$. The second-best fit was 20.15% for $Pnm'a$ (#62.444), which involves a combination of $\psi_8$ and $\psi_9$ within $\Gamma_6$. All other MSGs resulted in $R_{mag}$ values exceeding 25%. We anticipate that the transition at $T_N$ is of second order, and according to Landau's theory,[39] only one irrep, *i.e.*, MSG, can be involved. The magnetic structure at 64.8(1) K corresponding to the best fit is illustrated in Fig. 1(b), with the estimated moment being 3.64(4)$\mu_B$ per Fe$^{3+}$ site. Magnetic moments aligned parallel to the $b$-axis form a simple antiferromagnetic structure, with the magnetic unit cell equivalent to the crystallographic unit cell. The magnetic moment size is traced as a function of temperature [Fig. 2(c)], with saturation nearly achieved at the lowest measured temperature. This explicit, group-theoretical, data-driven selection uniquely identifies $Pnma'$ (#62.445) from powder data and provides a reusable template for similar systems. To compare with previous work,[13] we note that the magnetic structure of α-FeOOH was previously reported in the alternative $Pbnm$ setting, with spins parallel to the $c$-axis. Under the axis transformation between the $Pbnm$ and $Pnma$ settings, this spin direction is equivalent to the $b$-axis moment direction obtained here. Thus, our refined magnetic structure is fully consistent with the previous neutron diffraction result.

In the paramagnetic phase of 430.1(9) K, diffraction patterns are well fit based on α-FeOOH with the space group $Pnma$. The refined crystal structure is consistent with the averaged structure obtained from single-crystal X-ray diffraction data.[40] The standard Rietveld refinements converge and successfully estimate the OH (O2–H) distance as a function of temperature [Fig. 2(d)]. The distance is close to, yet slightly larger than 0.96 Å in water, and it remains within a 5% variation across the entire temperature range measured. No abrupt change or anomaly is observed that would indicate either a proton order–disorder transition or a symmetry change of the average crystal structure. In addition to the O2–H bond length, we evaluated the H⋯O1 distance, the O1⋯O2 donor–acceptor distance, and the O2–H⋯O1 angle as a function of temperature (Fig. S1) to visualise the evolution of the hydrogen-bond geometry relevant to surface proton transfer.

Table 1 Basis vectors (BVs) of the irreducible representations (irreps) for the space group *Pnma* with the magnetic wavevector, $q_m$ = (0, 0, 0) r.l.u. with schematic drawings of the magnetic structures corresponding to each BV. The atoms are defined as follows: #1: (*x*, 1/4, *z*), #2: (−*x* + 1/2, 3/4, *z* + 1/2), #3: (−*x*, 3/4, −*z*), and #4: (*x* + 1/2, 1/4, −*z* + 1/2)

| Irrep | BV | Atom #1 | | | Atom #2 | | | Atom #3 | | | Atom #4 | | | |
|---|---|---|---|---|---|---|---|---|---|---|---|---|---|---|
| | | $m_x$ | $m_y$ | $m_z$ | $m_x$ | $m_y$ | $m_z$ | $m_x$ | $m_y$ | $m_z$ | $m_x$ | $m_y$ | $m_z$ | |
| $\Gamma_1$ | $\Psi_1$ | 0 | 2 | 0 | 0 | −2 | 0 | 0 | 2 | 0 | 0 | −2 | 0 | 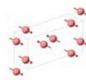 |
| $\Gamma_2$ | $\Psi_2$ | 2 | 0 | 0 | −2 | 0 | 0 | −2 | 0 | 0 | 2 | 0 | 0 | 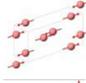 |
| | $\Psi_3$ | 0 | 0 | 2 | 0 | 0 | 2 | 0 | 0 | −2 | 0 | 0 | −2 | 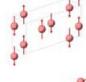 |
| $\Gamma_3$ | $\Psi_4$ | 2 | 0 | 0 | 2 | 0 | 0 | 2 | 0 | 0 | 2 | 0 | 0 | 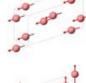 |
| | $\Psi_5$ | 0 | 0 | 2 | 0 | 0 | −2 | 0 | 0 | 2 | 0 | 0 | −2 | 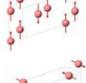 |
| $\Gamma_4$ | $\Psi_6$ | 0 | 2 | 0 | 0 | 2 | 0 | 0 | −2 | 0 | 0 | −2 | 0 | 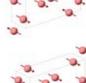 |
| $\Gamma_5$ | $\Psi_7$ | 0 | 2 | 0 | 0 | 2 | 0 | 0 | 2 | 0 | 0 | 2 | 0 | 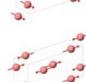 |
| $\Gamma_6$ | $\Psi_8$ | 2 | 0 | 0 | 2 | 0 | 0 | −2 | 0 | 0 | −2 | 0 | 0 | 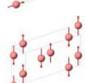 |
| | $\Psi_9$ | 0 | 0 | 2 | 0 | 0 | −2 | 0 | 0 | −2 | 0 | 0 | 2 | 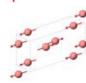 |
| $\Gamma_7$ | $\Psi_{10}$ | 2 | 0 | 0 | −2 | 0 | 0 | 2 | 0 | 0 | −2 | 0 | 0 | 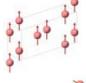 |
| | $\Psi_{11}$ | 0 | 0 | 2 | 0 | 0 | 2 | 0 | 0 | 2 | 0 | 0 | 2 | 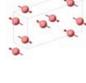 |
| $\Gamma_8$ | $\Psi_{12}$ | 0 | 2 | 0 | 0 | −2 | 0 | 0 | −2 | 0 | 0 | 2 | 0 | 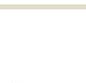 |

Although the hydrogen positions have been successfully determined, as also supported by the difference-Fourier map (Fig. S2), some concern may remain because an undeuterated (light hydrogen) sample was employed. The neutron-diffraction intensity *I* is proportional to the square of the absolute value of the structure factor, as formulated below,

$$I \propto \left| \sum_d b_d e^{2\pi i(hx_d + ky_d + lz_d)} e^{-W_d} \right|^2,$$

where $b_d$ stands for the coherent scattering length for the atom *d* with coordinates ($x_d$, $y_d$, $z_d$), *h*, *k*, *l* are Miller indices, $e^{-W_d}$ is the Debye–Waller factor, and the summation is over all atoms in the unit cell. The coherent scattering lengths for each element are Fe (9.45 fm), O (5.803 fm), and H (−3.7390 fm).[34] Hydrogen has an unusually negative scattering length, whereas the others are positive. Due to this negative scattering length, hydrogen tends to cancel out the contributions of other elements in the intensity. Since the other elements have positive scattering lengths, combining these values results in a complex variation in the overall intensity. Simulated diffraction patterns are plotted for cases with only hydrogen [Fig. 3(b)] and without hydrogen [Fig. 3(c)], highlighting the effects of hydrogen presence. Although several low-angle reflections are significantly influenced by hydrogen content, complementary first-principles calculations should be performed to further confirm the hydrogen positions.

Table 2 The general positions of the maximal magnetic space groups (MSGs) of the parent space group Pnma, with the magnetic wavevector $q_m$ = (0, 0, 0) r.l.u. The unified (UNI) symbols[38] for MSGs are employed. Corresponding irreducible representations are also shown

| MSG | Fe positions | Magnetic moment | Corr. irrep |
| --- | --- | --- | --- |
| Pn'm'a' (#62.449) | $(x, 1/4, z\|m_x, 0, m_z)$ $(-x + 1/2, 3/4, z + 1/2\|-m_x, 0, m_z)$ $(-x, 3/4, -z\|-m_x, 0, -m_z)$ $(x + 1/2, 1/4, -z + 1/2\|m_x, 0, -m_z)$ | $(M_x, 0, M_z)$ | $(\psi_2 \oplus \psi_3) \in \Gamma_2$ |
| Pn'ma' (#62.448) | $(x, 1/4, z\|0, m_y, 0)$ $(-x + 1/2, 3/4, z + 1/2\|0, m_y, 0)$ $(-x, 3/4, -z\|0, m_y, 0)$ $(x + 1/2, 1/4, -z + 1/2\|0, m_y, 0)$ | $(0, M_y, 0)$ | $\psi_7 \in \Gamma_5$ |
| Pnm'a' (#62.447) | $(x, 1/4, z\|m_x, 0, m_z)$ $(-x + 1/2, 3/4, z + 1/2\|m_x, 0, -m_z)$ $(-x, 3/4, -z\|m_x, 0, m_z)$ $(x + 1/2, 1/4, -z + 1/2\|m_x, 0, -m_z)$ | $(M_x, 0, M_z)$ | $(\psi_4 \oplus \psi_5) \in \Gamma_3$ |
| Pn'm'a (#62.446) | $(x, 1/4, z\|m_x, 0, m_z)$ $(-x + 1/2, 3/4, z + 1/2\|-m_x, 0, m_z)$ $(-x, 3/4, -z\|m_x, 0, m_z)$ $(x + 1/2, 1/4, -z + 1/2\|-m_x, 0, m_z)$ | $(M_x, 0, M_z)$ | $(\psi_{10} \oplus \psi_{11}) \in \Gamma_7$ |
| Pnma' (#62.445) | $(x, 1/4, z\|0, m_y, 0)$ $(-x + 1/2, 3/4, z + 1/2\|0, -m_y, 0)$ $(-x, 3/4, -z\|0, -m_y, 0)$ $(x + 1/2, 1/4, -z + 1/2\|0, m_y, 0)$ | $(0, M_y, 0)$ | $\psi_{12} \in \Gamma_8$ |
| Pnm'a (#62.444) | $(x, 1/4, z\|m_x, 0, m_z)$ $(-x + 1/2, 3/4, z + 1/2\|m_x, 0, -m_z)$ $(-x, 3/4, -z\|-m_x, 0, -m_z)$ $(x + 1/2, 1/4, -z + 1/2\|-m_x, 0, m_z)$ | $(M_x, 0, M_z)$ | $(\psi_8 \oplus \psi_9) \in \Gamma_6$ |
| Pn'ma (#62.443) | $(x, 1/4, z\|0, m_y, 0)$ $(-x + 1/2, 3/4, z + 1/2\|0, m_y, 0)$ $(-x, 3/4, -z\|0, -m_y, 0)$ $(x + 1/2, 1/4, -z + 1/2\|0, -m_y, 0)$ | $(0, M_y, 0)$ | $\psi_6 \in \Gamma_4$ |
| Pnma.1 (#62.441) | $(x, 1/4, z\|0, m_y, 0)$ $(-x + 1/2, 3/4, z + 1/2\|0, -m_y, 0)$ $(-x, 3/4, -z\|0, m_y, 0)$ $(x + 1/2, 1/4, -z + 1/2\|0, -m_y, 0)$ | $(0, M_y, 0)$ | $\psi_1 \in \Gamma_1$ |

Fig. 4 shows the calculated distribution of magnetisation density ($\Delta\rho_{\uparrow\downarrow}$) for α-FeOOH, determined using the most stable antiferromagnetic configuration from a series of total energy calculations. This antiferromagnetic state is 0.038 eV per atom, more stable than the ferromagnetic state. $\Delta\rho_{\uparrow\downarrow}$ is calculated from the charge densities of up-spin ($\rho_\uparrow$) and down-spin ($\rho_\downarrow$) using the equation,

$$\Delta\rho_{\uparrow\downarrow} = \rho_\uparrow - \rho_\downarrow.$$

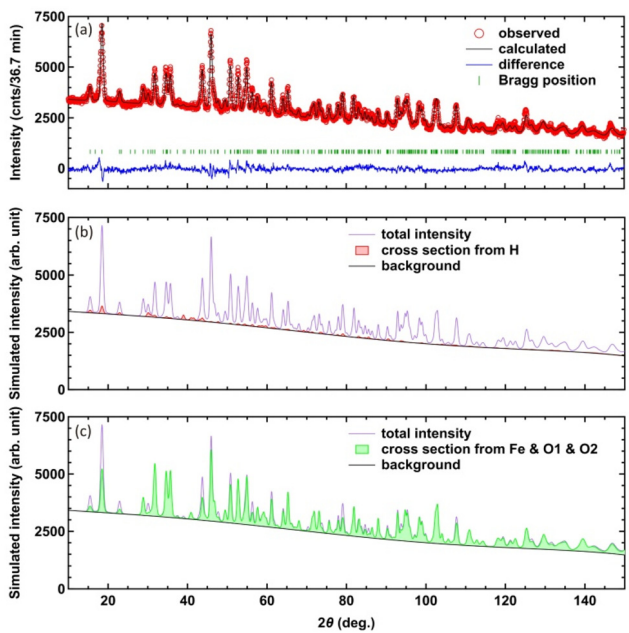

Fig. 3 (a) Neutron powder diffraction data for 430.1(9) K with Rietveld refinement (solid lines). The calculated positions of nuclear and magnetic reflections are indicated by green ticks. The bottom lines show the difference between observed and calculated intensities. Simulated intensities are plotted for (b) the case with only hydrogen and (c) without hydrogen.

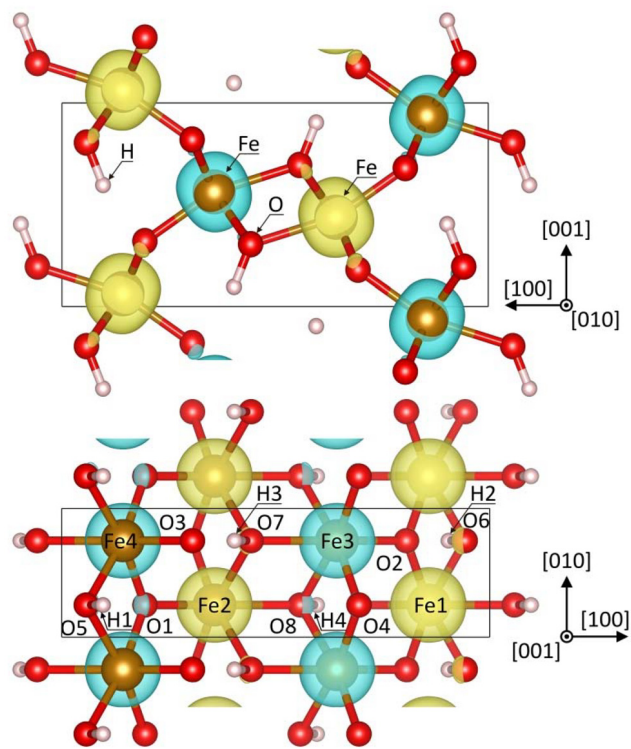

Fig. 4 The spatial distribution of magnetisation density in α-FeOOH with its most stable antiferromagnetic configurations. The yellow and light-blue isosurfaces represent values of 0.2 and −0.2 Å$^{-3}$, respectively.

Table 3 Comparison of the crystal structure parameters of α-FeOOH between experimental and calculated data, and the magnetic structure under magnetic space groups (MSGs), with basic information on the relationships to the parent paramagnetic structure. Notations for Belov–Neronova–Smirnova (BNS),[44] Opechowski–Guccione (OG),[45] and unified (UNI)[38] symbols are employed

|  | Experiment | Calculation |
|---|---|---|
| Temperature | 64.8(1) K |  |
| Parent space group | *Pnma* (#62) |  |
| Magnetic wavevector | (0, 0, 0) r.l.u. |  |
| Transformation from the parent basis | (***a***, ***b***, ***c***; 0, 0, 0) |  |
| MSG symbol | BNS: *Pnma*′ (#62.445) |  |
|  | OG: *Pnma*′ (#62.5.506) |  |
|  | UNI: *Pnma*′ (#62.445) |  |
| Transformation to the standard setting | (***a***, ***b***, ***c***; 0, 0, 0) |  |
| Magnetic point group[43] | *mmm*′ |  |
| Unit cell parameters | $a$ = 9.93455(3) Å | $a$ = 9.88511 Å |
|  | $b$ = 3.01209(8) Å | $b$ = 3.01004 Å |
|  | $c$ = 4.59219(13) Å | $c$ = 4.54221 Å |
| MSG symmetry operations | $x, y, z, +1$ $\{1|0,0,0\}$ |  |
|  | $-x + 1/2, -y, z + 1/2, +1$ $\{2_{001}|1/2, 0, 1/2\}$ |  |
|  | $-x + 1/2, y + 1/2, z + 1/2, +1$ $\{m_{100}|1/2, 1/2, 1/2\}$ |  |
|  | $x, -y + 1/2, z, +1$ $\{m_{010}|0, 1/2, 0\}$ |  |
| MSG symmetry centering operations | $x, y, z, +1$ $\{1|0, 0, 0\}$ |  |
|  | $-x, -y, -z, -1$ $\{-1'|0, 0, 0\}$ |  |
| Positions of magnetic atoms | Fe (0.85271(19), 1/4, 0.04986(42)) | Fe (0.85715, 1/4, 0.06233) |
| Magnetic moments components | $M_y$ = 3.64(4) $\mu_B$ |  |
| Positions of nonmagnetic atoms | O1 (0.20146(34), 1/4, 0.70280(69)) | O1 (0.19457, 1/4, 0.67982) |
|  | O2 (0.05209(31), 1/4, 0.20266(70)) | O2 (0.05748, 1/4, 0.19500) |
|  | H (0.08306(63), 1/4, 0.41172(138)) | H (0.09636, 1/4, 0.40360) |
| Isotropic displacement parameters, $U_{iso}$ (Å²) | Fe: 0.0040(4) | N/A |
|  | O1: 0.0013(6) |  |
|  | O2: 0.0028(6) |  |
|  | H: 0.0172(11) |  |

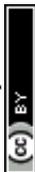

As shown in Fig. 4, the spin configurations of Fe–Fe are antiferromagnetic along the [100] direction and ferromagnetic along the [010] direction, which is consistent with the refined magnetic structure.

The optimised lattice constants are summarised in Table 3, along with the refined nuclear and magnetic structures. The experimentally refined parameters in Table 3 are described within the MSG in accordance with the latest guidelines of the International Union of Crystallography,[41,42] including the positions of nonmagnetic atoms. The lattice constants obtained from our calculations are $a$ = 9.885, $b$ = 3.010, and $c$ = 4.542 Å. The relative differences between the computational lattice constants and the experimental findings at 64.8(1) K were below 1%. Using Debye-Grüneisen fits to the lattice parameters as a function of temperature (Fig. S3), we extrapolated the experimental cell to 0 K and obtained $a$ = 9.93432 Å, $b$ = 3.01191 Å, and $c$ = 4.59270 Å, which are marginally closer to the calculated lattice constants than the 64.8(1) K values. The fractional atomic positions in the optimised α-FeOOH closely match the experimental results, as summarised in Table 3. The O2–H distance is computed to be 1.02253 Å compared to the experimentally refined value of 1.008(7) Å, within an accuracy of 1.5%. To quantify consistency at the level of hydrogen bonding—not only lattice constants—we evaluated the donor–acceptor geometry using the refined/optimised fractional coordinates and lattice parameters in Table 3. For 64.8(1) K neutron refinement, the geometry is H⋯O1 = 1.781(7) Å, O1⋯O2 = 2.734(4) Å, and ∠O2–H⋯O1 = 156.4(6)°. For the DFT+U-optimised structure, the corresponding values are H⋯O1 = 1.586 Å, O1⋯O2 = 2.704 Å, and ∠O2–H⋯O1 = 164.3°, giving the same donor–acceptor pairing and a nearly linear hydrogen bond arrangement.

Surface slab models of α-FeOOH are commonly built from the bulk structure, and the initial protonation pattern of surface O sites is inherited from the bulk OH orientation and hydrogen-bond topology. If the bulk H position is poorly constrained, multiple plausible hydrogen-bond networks and surface hydroxyl configurations must be assumed, which can lead to different predicted adsorption geometries and proton-coupled electron transfer energetics for $CO_2$ reduction. The present refinement provides a well-defined bulk reference (including temperature-dependent hydrogen-bond geometry) that can be used as a consistent starting point for reaction modelling and for assessing how defects or surface terminations perturb the proton environment.

Fig. 5 shows the calculated projected partial density of states (PDOS) for α-FeOOH. The valence band top is set to 0 eV on the horizontal axis. The band gap of antiferromagnetic α-FeOOH is calculated to be 2.16 eV. The regions around the valence band top and the conduction band bottom are composed of Fe(3d) and O(2p) orbitals. Unlike O1–4, O5–O8 [Fig. 4] are combined with a proton to form OH groups. The combinations of atoms forming OH ions are O6–H2, O7–H3, O5–H1, and O8–H4. The bonding states of H(1s)–O(2p) and H(1s)–O(2s) are present around −5 to −8 eV and −18 to −20 eV, respectively. The molecular orbitals of OH are located at deeper valence bands than the energy levels of the O(2p) states from the single oxide ions, O1–O4.

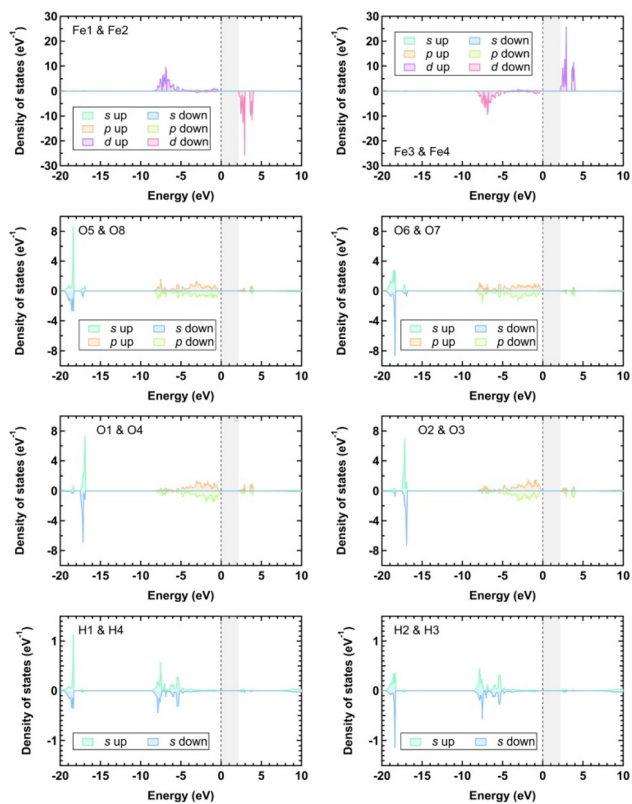

**Fig. 5** Calculated partial density of states (PDOSs) of α-FeOOH. Atom labels correspond to those in Fig. 4. The energy level of the valence band top is set to be 0 eV in the horizontal axis. The hatched gray area indicates the band gap region. Positive and negative PDOSs correspond to up-spin and down-spin components, respectively.

Given the nearly complete agreement between experimental and calculated data, we show that the complementary use of neutron diffraction on an undeuterated sample and first-principles calculations can effectively identify hydrogen positions. Due to the high symmetry of this particular compound, refinements were relatively straightforward. The results from both experiments and calculations were highly consistent, highlighting the efficacy of a parallel approach using neutrons and calculations. Deuteration may be necessary for compounds with more complex structures and numerous Wyckoff positions. Hydrogen positions in α-FeOOH have been successfully refined, providing the minimum information required to elucidate the mechanism of $CO_2$ reduction and water oxidation. Future perspectives include excitation studies on hydrogen dynamics and *in situ* measurements under a $CO_2$ atmosphere.

## Conclusions

In summary, we have refined the magnetic structure and hydrogen positions for goethite α-FeOOH. The magnetic structure was successfully refined using a powder sample under zero field based upon irreducible representation and magnetic space group analyses. Hydrogen positions were accurately determined through Rietveld refinements and first-principles calculations. The consistency between experimental and calculated results underscores the effectiveness of using both methods, even for undeuterated samples.

## Author contributions

Y. N. and K. M. conceived the project idea. Y. N. and M. K. conducted the neutron diffraction measurements, with Y. N. analysing the experimental data. A. K. performed the first-principles calculations. Y. N. and A. K. prepared the figures, while Y. N., A. K., and K. M. wrote the manuscript. All authors reviewed the manuscript.

## Conflicts of interest

The authors declare no competing interests.

## Data availability

The neutron diffraction data and refinement files are available from the corresponding author upon reasonable request.

Supplementary information (SI) is available. See DOI: https://doi.org/10.1039/d5dt00217f.

CCDC 2535200–2535209 contain the supplementary crystallographic data for this paper.[46a–j]

## Acknowledgements

We thank T. Arima and G. J. Nilsen for their valuable discussions. This work was supported by the JSPS (No. JP21H03732, JP22H05145, JP24K00572, JP22H05146, JP22H05142, JP22H05148), FOREST (No. JPMJFR202V) from JST, "Program for Promoting Researches on the Supercomputer Fugaku" (JPMXP1020230325) from MEXT, and the Graduate Program in Spintronics at Tohoku University. Work on HERMES at JRR-3 was performed using the instrument research team beamtime (proposal no. 22410).